\documentclass[12pt,preprint]{aastex}

\usepackage{epsfig}
\usepackage{emulateapj5}
\usepackage{apjfonts}

\newcommand{\chandra}{{\it Chandra}}

\newcommand{\xmm}{{\it XMM-Newton}}

\newcommand{\lum}{\thinspace\hbox{$\hbox{ergs}\thinspace\hbox{s}^{-1}$}}
\newcommand{\flux}{\thinspace\hbox{$\hbox{ergs}\thinspace\hbox{cm}^{-2}\thinspace\hbox{s}^{-1}$}}
\newcommand{\hst}{{\it HST}}

\makeatletter

\newenvironment{inlinefigure}{%
\def\@captype{figure}%
\noindent\begin{minipage}{0.999\linewidth}\begin{center}}
{\end{center}\end{minipage}\smallskip}

\begin{document}

\def\spose#1{\hbox to 0pt{#1\hss}}
\def\laeq{\mathrel{\spose{\lower 3pt\hbox{$\mathchar"218$}}
     \raise 2.0pt\hbox{$\mathchar"13C$}}}
\def\gaeq{\mathrel{\spose{\lower 3pt\hbox{$\mathchar"218$}}
     \raise 2.0pt\hbox{$\mathchar"13E$}}}

\slugcomment{Accepted for publication in ApJ}

\title{X-ray Localization of the Globular Cluster G1 with {\it 
XMM-Newton}} 

\author{Albert~K.~H.~Kong}
\affil{Kavli Institute for Astrophysics and Space Research,
Massachusetts Institute of Technology, 77
Massachusetts Avenue, Cambridge, MA 02139; akong@space.mit.edu}

\begin{abstract}
We present an accurate X-ray position of the massive globular cluster G1 
by using \xmm\ and the {\it Hubble Space Telescope} ({\it HST}). The 
X-ray emission of G1 has been detected 
recently with \xmm. There are two possibilities for the origin of 
the X-ray emission. It can be either due to accretion of the central 
intermediate-mass black hole, or by ordinary low-mass X-ray binaries. The 
precise location of the X-ray emission might distinguish between these two 
scenarios. By refining the astrometry of the \xmm\ and \hst\ data, we 
reduced the \xmm\ error circle to $1.5''$. Despite the smaller error circle, the precision is not sufficient to distinguish an intermediate-mass black hole and luminous low-mass X-ray binaries.
This result, however, suggests that future \chandra\ observations may 
reveal the origin of the X-ray emission. 
\end{abstract}

\keywords{binaries: close---globular clusters: individual 
(Mayall II = G1)---X-rays: binaries}

\section{Introduction}
Globular clusters are very efficient places to produce X-ray binaries 
via dynamical interactions. In particular, it has been known for 
many years that the formation rate per unit mass of
luminous ($L_X >10^{36}$\lum) X-ray sources is 
much higher in globular clusters than that of the rest of our Galaxy. 
More recently, similar 
results are found in other nearby spiral galaxies like M31 (Di\,Stefano 
et al. 2002) and M104 (Di\,Stefano et al. 2003). Among all extragalactic 
globular clusters,  
G1 in M31 is an intriguing one. With a luminosity of $\sim 10^6 
L_\odot$ (Rich et al. 
1996), it is the
most luminous star cluster in the Local Group, and also one of the most 
massive, at $(7-17) \times 10^6 M_\odot$ (Meylan et al. 2001).
The rates
at which X-ray binaries are created in the cluster core are therefore 
expected
to be high compared with globular clusters in the Milky Way. 
Furthermore, it has
been claimed, based on kinematic studies, that G1 hosts a $\sim 2\times 
10^4 \, M_\odot$
intermediate-mass black hole (Gebhardt et al. 2002,2005). 
However, this result is controversial and has been challenged by 
Baumgardt et al. (2003). X-ray observations of G1 therefore allow us to 
investigate some of the interesting properties of the cluster.

Recently, {\it XMM-Newton} has conducted
three short ($<$ 10 ksec) observations of G1 and has discovered an
X-ray source coincident with G1 (Trudolyubov \& Priedhorsky 2004; Pooley 
\& Rappaport 2006). To explain the origin of the X-ray emission of G1, 
Pooley \& Rappaport (2006) proposed that it could be due to accretion of 
ionized cluster gas by a
central intermediate-mass black hole or it could be produced by a 
conventional X-ray
binary, and it is also possible to 
distinguish these two
scenarios by obtaining a precise localization of the X-ray
emission. If the X-ray emission is due to a central 20,000 $M_{\odot}$
black hole, we expect it comes from within 50 mas of the 
center. However, if low-mass X-ray
binaries are responsible to the X-ray emission, then we expect the X-ray emission offsets from the core. This requires high-resolution X-ray observations. However, there is no \chandra\ observation of G1 
and only \xmm\ observations are available. Although Pooley \& Rappaport 
(2006) investigated the \xmm\ spectra 
of G1 in detail, they did not perform an astrometric study. The absolute astrometry of \xmm\ is about $2''$ (Kirsch 2006) while the statistical uncertainty is intensity dependent. This leads to a positional error of about $2''-6''$ depending on the source brightness.
While 
the spatial resolution of {\it XMM-Newton} is much poorer than 
\chandra\, it is possible to localize X-ray positions to $1''-2''$ 
with \xmm\ if one calibrates the astrometry carefully. In this paper, 
we refined the X-ray position of G1 by performing precise relative 
astrometry using \xmm\ and the {\it Hubble Space Telescope} ({\it HST}).

\section{Observations and Data Analysis}
\subsection{\xmm}
G1 was first observed with the \xmm\ in 2001 January for a total 
exposure time of $\sim 8$ ksec. There were two more 
\xmm\ observations in 2002 December and 2003 February. Both observations were off-axis resulting heavy vignetting and 
one of the observations was affected by high background [see Pooley \& 
Rappaport (2006) for a summary]. In this 
Letter, we only consider the first observation taken in 2001. 
All three cameras (one pn camera and two MOS cameras) of the European 
Photon Imaging 
Camera (EPIC) were turned on for collecting data. 
All the X-ray
data were processed with the \xmm\ Science Analysis System (SAS)
version 7.0. 

We downloaded the raw data from the \xmm\ archive and 
reprocessed with SAS together with the latest calibration products. The 
reprocessed event lists were first examined for background variation 
using the high energy (10--15 keV) background lightcurves and we 
did not find any significant background flaring event. We extracted 
X-ray images with photon energies in the range of 0.3--10 keV, and 
only considered events with 
FLAG = 0 and single and double events for the pn camera (PATTERN $\leq 
4$), and single to quadruple events for the MOS cameras (PATTERN $\leq 
12$). Source detection was then performed using a maximum likelihood 
approach as implemented by the SAS tools {\it edetect\_chain}. We ran the 
source detection simultaneously on the data from all three cameras. G1 
was clearly detected and was seen in all three cameras with a combined 
detection likelihood of 63. We 
compared the X-ray source list with the 2MASS and USNO catalogs and 
images, and looked for coincidence of bright and isolated stellar 
objects. We found one star (2MASS\,00325251+3931424) that is $< 3''$ from 
the X-ray position and it is likely to be a foreground star. 

To verify the nature of this X-ray 
emitting stellar object, we computed the hardness ratios. These ratios 
were based 
on the source counts in the three energy bands: $S$ (0.3--1 keV), $M$ 
(1--2 keV), and $H$ (2-10 keV). The two hardness ratios are defined as 
HR1=$(M-S)/(M+S)$ and HR2=$(H-S)/(H+S)$.  Figure 1 shows the color-color 
diagram of the X-ray emitting foreground star and G1. We have overlaid 
the color-color diagram with lines showing the tracks followed by 
representative spectra with differing values of $N_H$.
The X-ray colors of the X-ray emitting star indicate that it 
has a very soft X-ray spectrum, consistent with a very soft X-ray source 
(Di\,Stefano \& Kong 2004).
The X-ray radiation is therefore likely due to the coronal 
emission from a foreground star. The star has a $R$ magnitude of 14.0 
(Monet et al. 2003). We calculated the X-ray to optical flux ratio as 
$\log(f_X/f_R)=\log f_X + 5.67 + 0.4 R$ (Hornschemeier et al. 2001). 
With a count rate of 0.018 c/s in the pn detector and assuming a 
Raymond-Smith model with 
$kT_{RS}=0.3$ keV and $N_H=10^{21}$ cm$^{-2}$, the 0.3--10 keV flux is 
$2.8\times10^{-14}$\flux and the 
corresponding $f_X/f_R$ is 0.005, typical for a foreground star 
(Hornschemeier et al. 2001). 

Based on the optical counterpart, the boresight 
correction that needs to be applied to the X-ray source positions is 
$1.47''\pm0.86''$ in R.A. and $2.51''\pm0.86''$ in decl.; the uncertainties are quadratic sum of the positional errors of the X-ray and 2MASS source. The correction is 
consistent with 
the absolute pointing accuracy of \xmm\ (see also an example in Pietsch, 
Freyberg \& Haberl 2005).

\vspace{2mm}
\begin{inlinefigure}
\psfig{file=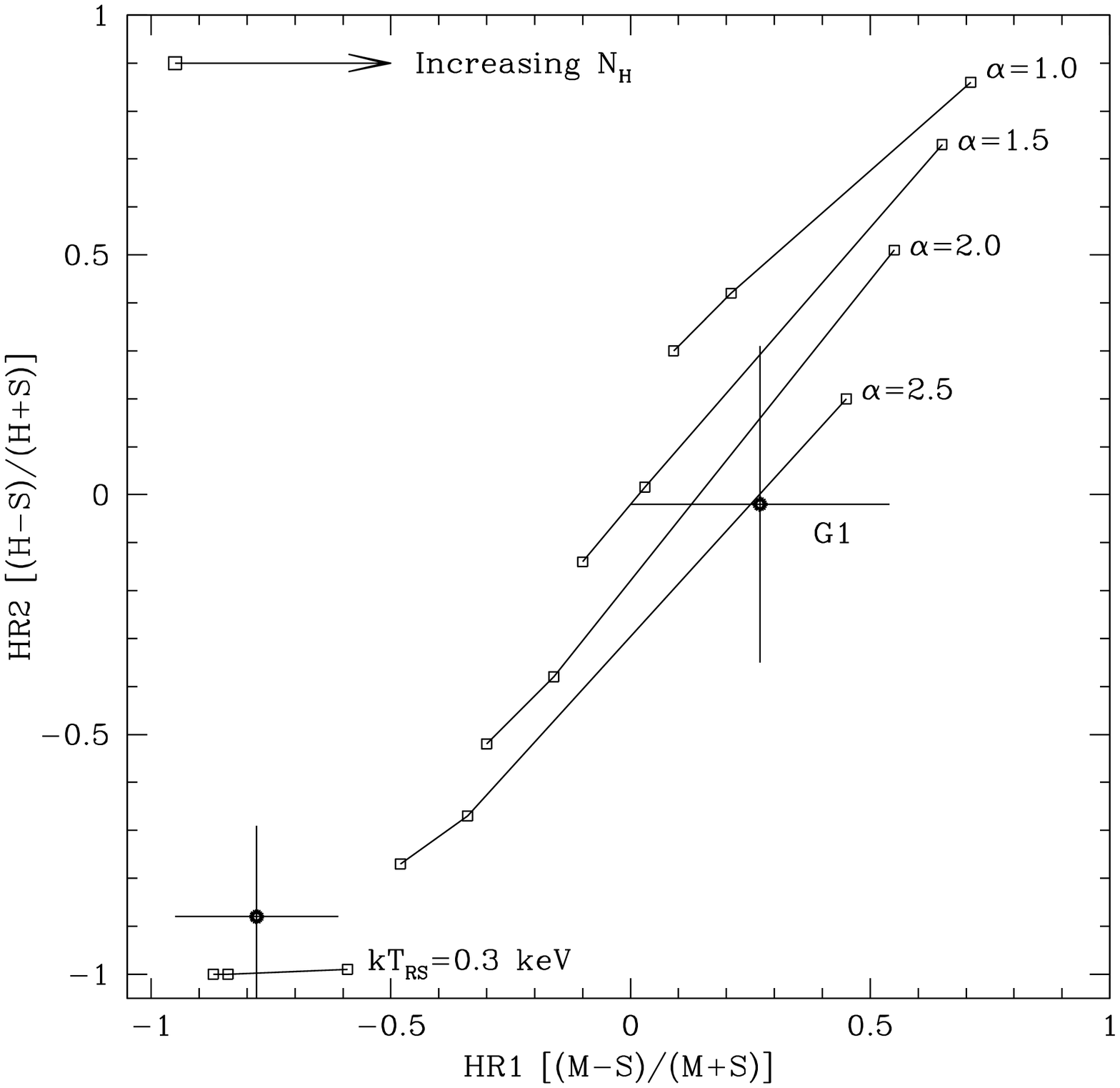,width=3.4in}
\caption{Color-color diagram of G1 and a nearby X-ray emitting object 
(lower left).
Also plotted are the estimated hardness ratios estimated from different 
spectral models. Top to bottom: Power-law model with $\alpha$ of 1.0, 
1.5, 2.0, and 2.5 and Raymond-Smith model with $kT_{RS}$ of 0.3 keV. 
For each model, $N_H$ 
varies from the left from $5\times10^{20}$, $10^{21}$, and 
$5\times10^{21}$ cm$^{-2}$.}
\end{inlinefigure}

\subsection{{\it Hubble Space Telescope}}
G1 was observed with the \hst\ Advanced Camera for Surveys (ACS) in High 
Resolution Channel (HRC) mode 
on 2003 October 24. The total integration time is 41 
minutes in the F555W filter centering on G1. We used the \hst\ pipeline 
data that were shifted and co-added using the MultiDrizzle package in 
PyRAF, with 
masking of cosmic rays, saturated pixels, and bad pixels.
We calibrated the astrometry of the \hst\ data by using the 2MASS 
catalog. The field-of-view of the ACS/HRC is small ($29''\times26''$) and only two stars in the field are in the 2MASS catalog. By computing the average offset between the \hst\ and 2MASS stars, we shifted the \hst\ image by $0.77634''$ in R.A. and $-0.5814''$ in decl., with a residual of $0.13''$. The ACS/HRC image of G1 is shown in Figure 2.

G1 was also observed with the \hst\ Wide Field Planetary Camera 2 (WFPC2) 
on 1995 October 2 with a total integration time of 37 minutes 
in the F555W filter. In addition to the F555W data, images were also taken in 
the F814W and F1042M filters. We downloaded the F555W image from the WFPC2 
Associations Science Products Pipeline for which cosmic-ray free, 
science-quality images are dithered and co-added. Since the field-of-view of 
WFPC2 is much larger than that of 
ACS/HRC, we can correct the astrometry with 6 stars in the 2MASS catalog. We applied the astrometry correction using IRAF task {\it ccmap} yielding a residual of $0.14''$. The 
WFPC2 F555W image is shown in Figure 2.

\section{X-ray Localization of G1}

After registering the absolute reference frames of the \hst\ and 
\xmm\ images to the 2MASS catalog, we 
located the center of G1 in the \hst\ images and the X-ray position of 
G1. We determined the centroid of G1 in the ACS/HRC image (R.A.=00h32m46.537s, 
decl.=+39d34m40.65s with $1\sigma$ error of $0.004''$) by computing the intensity weighted mean within the core radius ($0.21''$; Ma et al. 2007) using IRAF task {\it center}. We also checked the result by using the half-mass radius ($1.73''$) and there is no difference except for a larger error bar ($0.01''$). For the X-ray position, we applied 
the astrometric correction on the value determined by {\it edetect\_chain} 
yielding R.A.=00h32m46.6s, decl.=+39d34m40s. We 
then determined the $1\sigma$ radius 
error circle ($1.5''$) of the \xmm\ position of G1 by computing the 
quadratic sum 
of the positional uncertainty for the X-ray source ($1.23''$), the 
uncertainty in the optical astrometry ($0.13''$), and the uncertainty in 
the X-ray boresight correction ($0.86''$). The same procedure was also 
applied to the WFPC2 image. Figure 2 
shows the ACS/HRC and WFPC2 images 
of G1 and the $1\sigma$ radius X-ray error circles. The center 
of G1 derived from the optical image is also marked.

\begin{figure*}
\psfig{file=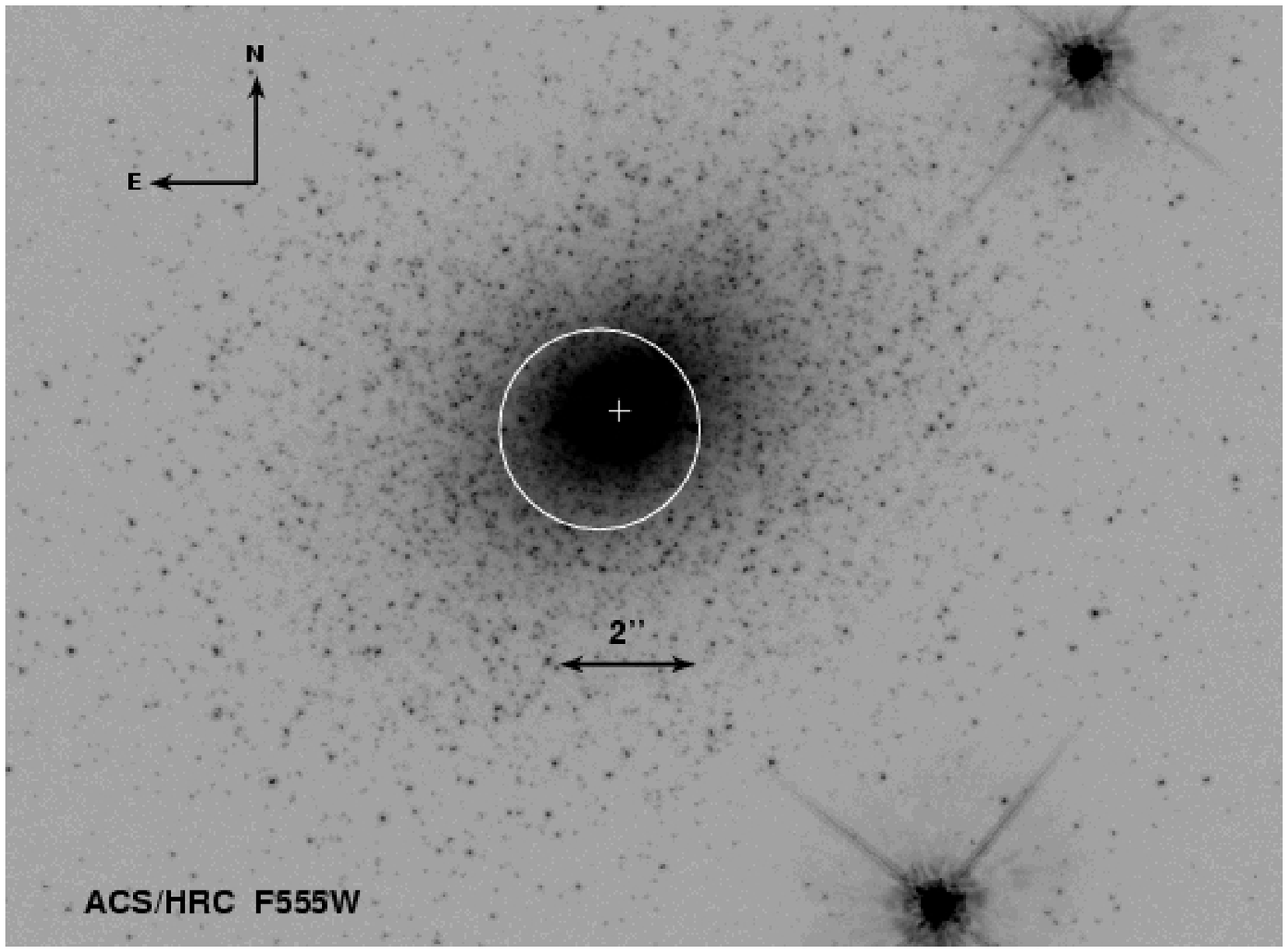,width=3.65in}
\psfig{file=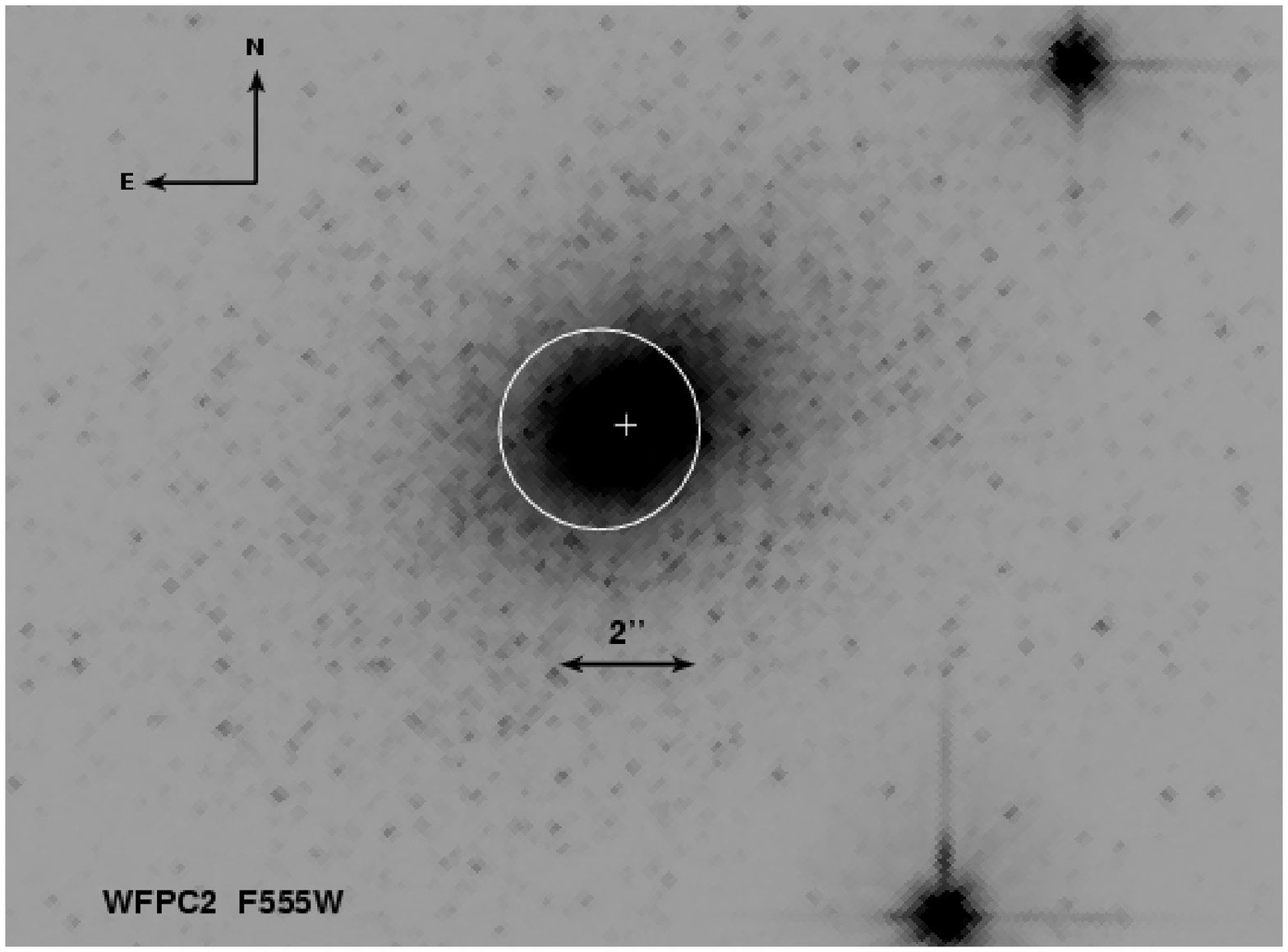,width=3.65in}
\caption{
\hst\ ACS/HRC F555W (left) and WFPC2 F555W (right) images of G1.
The cluster core is marked by a cross. The circles of both images are the $1\sigma$ radius ($1.5''$) \xmm\ error circles. The two bright stars in the field were used for calibrating the astrometry with the 2MASS catalog.}
\end{figure*}

\section{Discussion}
Using \xmm\ and \hst, we determined the centroid of G1 in the optical 
images as well as the X-ray position of G1. From Figure 2, although the 
X-ray position offsets from the cluster core,
the cluster center is within the $1\sigma$ error circle of the 
X-ray position. Therefore, the current \xmm\ data 
cannot provide 
constraint on whether the X-rays of G1
come from Bondi accretion of ionized cluster gas by a central 
intermediate-mass black hole for which the X-rays should come from the 
central 50 mas of the cluster (Pooley \& Rappaport 2006). Alternatively, 
the X-ray emission could be produced by luminous low-mass X-ray binaries 
 and we expect such X-ray emission may be outside the cluster 
center. Previous X-ray observations of globular clusters suggest that luminous ($L_X\gaeq10^{36}$ ergs s$^{-1}$) X-ray sources tend to locate within the core radius (Grindlay et al. 1984). Recent \chandra\ observations also show that nearly half of the quiescent low-mass X-ray binaries are found within the core radius (e.g., Grindlay et al. 2002; Pooley et al. 2002; Heinke et al. 2003,2006). Therefore, it is likely that a luminous low-mass X-ray binary would locate within the core radius ($0.21''$; Ma et al. 2007) of G1. The X-ray emission of G1 could also come from multiple low-mass X-ray binaries and we may be able to resolve G1 as an extended source with high spatial resolution instrument. Pooley \& Rappaport (2006) estimated that about 75 low-mass X-ray binaries might be in G1. It is worth noting that only one globular cluster, M15, is known to host two luminous X-ray sources (White \& Angelini 2001; 
Hannikainen et al. 2005). In conclusion, based on the current \xmm\ data, 
we cannot distinguish the two possible mechanisms of generating the 
X-ray emission of G1.

While the X-ray position of G1 is the most crucial factor to determine 
its nature, Pooley \& Rappaport (2006) also suggested that the X-ray 
spectrum may provide some hint. However, it has been proven that 
X-ray spectra of intermediate-mass black hole candidates consist of a 
class of different spectral shapes and in many cases, the spectra can be fit with several models and the estimated mass 
of the accreting black hole is model dependent (e.g. Stobbart et al. 2006; Gon{\c c}alves \& Soria 2006). Some intermediate-mass 
black hole candidates also show X-ray spectral change at different 
luminosity states (Kong \& Di\,Stefano 2005). These make interpretation based on X-ray spectra more difficult.
Nevertheless, if we assume a simple 
accretion disk model for G1, following Pooley \& Rappaport (2006), we 
would expect G1 has a 10 eV supersoft component (see Di\,Stefano \& 
Kong 2003); any emission above 0.5 keV must come from additional components. From the color-color diagram (Figure 1), G1 has significant 
emission above 1 keV and indeed it is very similar to a typical X-ray 
binary in M31 with a simple power-law spectral model (Kong et al. 2002). 
Therefore, if G1 has a 10 eV supersoft spectrum, it must also have an 
additional hard component. Indeed, it would be a challenge for \xmm\ and 
\chandra\ to detect such supersoft emission because an absorbed 
(Galactic value to the direction of M31; $N_H=7\times10^{20}$ cm$^{-2}$) 
10 eV spectrum turns over at about 0.2 keV which is the sensitivity 
limit of these instruments. For instance, if the X-ray emission is 
dominated by a 10 eV spectrum, simulation shows that it requires 700 
ksec \xmm\ or 1 Msec \chandra\ observing time in order to detect the source. If the black hole of G1 is only 100 $M_\odot$, the thermal emission would have a temperature of about 80 eV and we should be able to detect it with a 6 ksec \xmm\ or 10 ksec \chandra\ observaiton. As a supersoft X-ray source ($kT < 100$ eV), we do not expect to see X-rays above 1 keV which is not consistent with our current result. 
Alternatively, if the X-ray emission is from a luminous 
low-mass X-ray binary, we would also expect soft multi-color disk blackbody X-ray 
emission ($kT_{in}\approx0.3-3$ keV) in addition to a power-law like component associated with Comptonization of cooler photons (e.g. Sidoli et al. 2001). However, with only 
$\sim 70$ counts from all three \xmm\ detectors, we do not have a good 
constraint on the spectral model. In conclusion, X-ray spectra provided by \xmm\ and \chandra\ alone cannot provide convincing evidence for the nature of the X-ray emission from G1.

Although the X-ray position provided by \xmm\ cannot provide any 
reasonable constraint to the nature of the X-ray emission of G1, it 
suggests that future \chandra\ observations may resolve the problem. As 
discussed in Pooley \& Rappaport (2006), we can improve the relative astrometry 
of \chandra\ and \hst\ to $0.1''-0.2''$ if we can match a few \chandra\ sources to their optical counterparts (e.g. foreground stars or background active galactic nuclei). 
With such observations, we can accurately localize the X-ray emission of G1. However, the fact that a luminous low-mass X-ray binary is likely to locate $\laeq 0.21''$ from the cluster core suggests that we may not disentangle from the emission of a possible intermediate-mass black hole. Alternatively, a \chandra\ observation may be able to distinguish between multiple X-ray sources as an extended object and point-like emission.

\begin{acknowledgements}
We would like to thank an anonymous referee for useful comments.
This work is based on observations obtained with \xmm, an ESA mission
with instruments and contributions directly funded by ESA member
states and the US (NASA). The \hst\ data presented in this paper were 
obtained from the Multimission Archive at the Space Telescope Science Institute 
(MAST). STScI is operated by the Association of Universities for 
Research in Astronomy, Inc., under NASA contract NAS5-26555. 
\end{acknowledgements}

{\it Facilities:} \facility{XMM (EPIC)}, \facility{HST (ACS/HRC, WFPC2)}

\end{document}